\documentclass[conference]{IEEEtran}
\IEEEoverridecommandlockouts
\usepackage{cite}
\usepackage{amssymb, amsmath, amsthm, graphicx, enumerate}
\usepackage{multirow}
\usepackage{setspace}
\usepackage{algorithm}
\usepackage{algpseudocode}
\usepackage{array}
\usepackage{subfigure}
\usepackage{float}
\usepackage{url}
\usepackage{textcomp}

\def\BibTeX{{\rm B\kern-.05em{\sc i\kern-.025em b}\kern-.08em
    T\kern-.1667em\lower.7ex\hbox{E}\kern-.125emX}}
\begin{document}

\title{Semi-supervised Multimodal Hashing}

\author{\IEEEauthorblockN{Dayong Tian}
\IEEEauthorblockA{\textit{School of Electronics and Information} \\
\textit{Northwestern Polytechnical University}\\
Xi'an, China \\
dayong.tian@nwpu.edu.cn}
\and
\IEEEauthorblockN{Maoguo Gong}
\IEEEauthorblockA{\textit{Key Laboratory of Intelligent Perception and Image Understanding} \\
\textit{Xidian University}\\
Xi'an, China\\
gong@ieee.org}
\and
\IEEEauthorblockN{Deyun Zhou}
\IEEEauthorblockA{\textit{School of Electronics and Information} \\
\textit{Northwestern Polytechnical University}\\
Xi'an, China \\
dyzhou@nwpu.edu.cn}
\and
\IEEEauthorblockN{Jiao Shi}
\IEEEauthorblockA{\textit{School of Electronics and Information} \\
\textit{Northwestern Polytechnical University}\\
Xi'an, China \\
jiaoshi@nwpu.edu.cn}
\and
\IEEEauthorblockN{Yu Lei}
\IEEEauthorblockA{\textit{School of Electronics and Information} \\
\textit{Northwestern Polytechnical University}\\
Xi'an, China \\
leiy@nwpu.edu.cn}
}

\maketitle

\begin{abstract}
Retrieving nearest neighbors across correlated data in multiple modalities, such as image-text pairs on Facebook and video-tag pairs on YouTube, has become a challenging task due to the huge amount of data. Multimodal hashing methods that embed data into binary codes can boost the retrieving speed and reduce storage requirement. As unsupervised multimodal hashing methods are usually inferior to supervised ones, while the supervised ones requires too much manually labeled data,  the proposed method in this paper utilizes a part of labels to design a semi-supervised multimodal hashing method. It first computes the transformation matrices for data matrices and label matrix. Then, with these transformation matrices, fuzzy logic is introduced to estimate a label matrix for unlabeled data. Finally, it uses the estimated label matrix to learn hashing functions for data in each modality to generate a unified binary code matrix. Experiments show that the proposed semi-supervised method with 50\% labels can get a medium performance among the compared supervised ones and achieve an approximate performance to the best supervised method with 90\% labels. With only 10\% labels, the proposed method can still compete with the worst compared supervised one.
\end{abstract}

\begin{IEEEkeywords}
information retrieval, multimodal hashing, fuzzy logic.
\end{IEEEkeywords}

\section{Introduction}
\IEEEPARstart{M}{ultimodal} data refer to correlated data of different types, such as image-text pairs in Facebook and video-tag pairs in Youtube. Multimodal hashing aims at embedding the multimodal data to binary codes in order to boost the speed of retrieval and reduce the storage requirement.\\
\indent Unsupervised multmodal hashing methods learn a hashing function to generate binary codes whose Hamming distances can ``simulate'' the Euclidean distances between each pair of data features. They assume the hand-crafted or learned features of images or texts can be linearly separated by hyperplanes. Hence, the Euclidean distances of data features in the same category are closer than those of data features in different categories. However, this assumption is impractical for large data sets of sophisticated data structures.\\
\indent Supervised multimodal hashing methods incorporate label information to improve the retrieval accuracy. With manually labeled data for training, these models are generally superior to unsupervised ones. However, they required a huge number of manually labeled data, which leads to a heavy burden on human experts. \\
\begin{figure*}[t]
\centering
\includegraphics[width=0.7\linewidth]{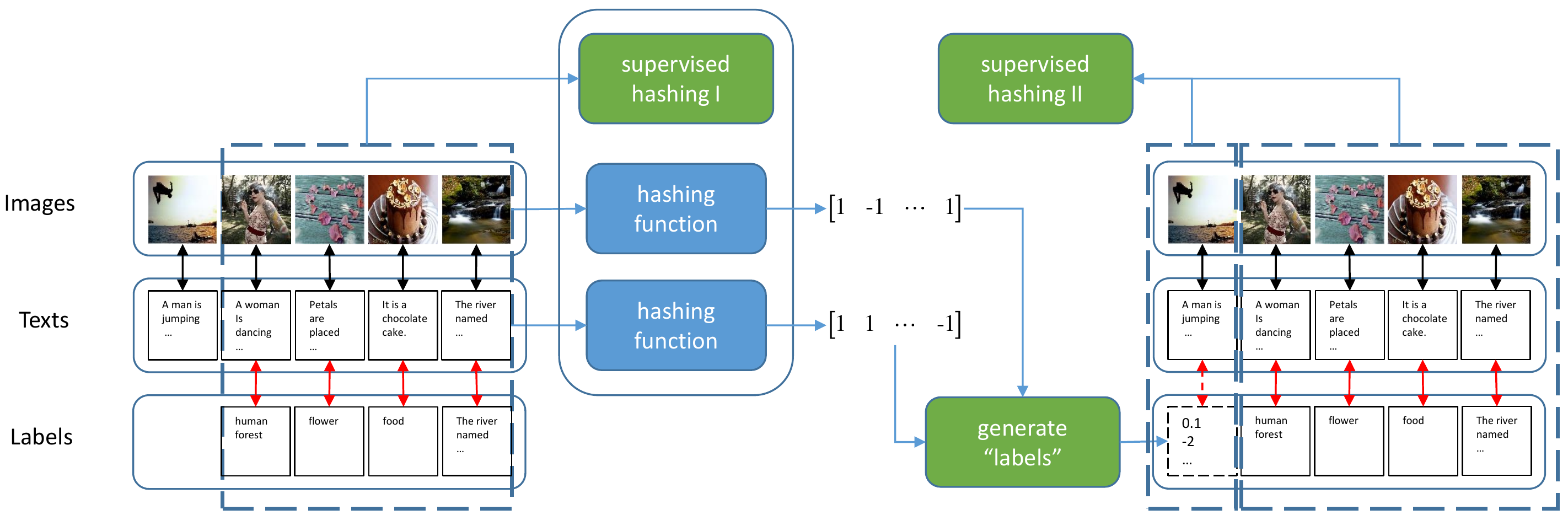}
\caption{Flowchart of SSMH. Supervised hashing I is for generating labels, while supervised hashing II is for generating hashing codes.}
\label{fig:scheme}
\end{figure*}
\indent To our best knowledge, semi-supervised semantic factorization hashing (S3FH)~\cite{S3FH} is the only multimodal hashing method. It generates a graph for each modality to avoid compute the pairwise distances in large data sets. Then the labels are estimated by a transformation of the unified hashing codes, which leads to incremental performance improvement as the number of available labels increases. In this paper, we proposed a semi-supervised multimodal hashing (SSMH) method by introducing fuzzy logic to estimate labels. The overall scheme of SSMH is given in Fig~\ref{fig:scheme}. SSMH first learns the hashing functions for different modalities of the labeled data. Note that labels which are represented by a binary matrix are treated as a special modality here. These hashing functions are used to generate candidate labels for the unlabeled but not the final hashing codes. Inspired by fuzzy c-means clustering, we introduce a membership variable to the hashing codes of unknown labels. Each membership variable represents the probability that the hashing code of an estimated label matches that of this modality. Finally, SSMH learns hashing functions to generate hashing codes.
\section{Related Works}
\subsection{Unimodal Hashing}
Unimodal hashing methods embeds a data matrix in one modality into binary code matrix. SH defines the unimodal hashing problem as:
\begin{equation}\label{eq:SH}
\begin{array}{c}
\mathop {\arg \min }\limits_{\bf{B}} \sum\limits_{{{\bf{x}}_i},{{\bf{x}}_j} \in {\bf{X}}} {{e^{ - {{\left\| {{{\bf{x}}_i} - {{\bf{x}}_j}} \right\|}^2}/{\sigma ^2}}}{{\left\| {{{\bf{b}}_i} - {{\bf{b}}_j}} \right\|}^2}} \\
s.t.\quad{\bf{B}} \in {\left\{ { - 1,1} \right\}^{n \times c}},\quad{{\bf{B}}^\top}{\bf{B}} = n{\bf{I}},\quad{{\bf{B}}^\top}{\bf{1}} = 0.
\end{array}
\end{equation}
where $\mathbf{X}$ is the data matrix of which each row is a data point, $\mathbf{x}_i$ is the $i$th row of $\mathbf{X}$, $\mathbf{B}$ is the hashing code matrix, $\mathbf{b}_i$ corresponds to the hashing code of $\mathbf{x}_i$, $n$ is the number of data points and $c$ is the code length. According to the inequality of arithmetic and geometric means, the object function of Eq.~\eqref{eq:SH} gets its minimum when
\begin{equation}
{{e^{ - {{\left\| {{{\bf{x}}_i} - {{\bf{x}}_j}} \right\|}^2}/{\sigma ^2}}}{{\left\| {{{\bf{b}}_i} - {{\bf{b}}_j}} \right\|}^2}}=constant,\quad \forall i\neq j.
\end{equation}
In this case, the Hamming distance of $\mathbf{b}_i$ and $\mathbf{b}_j$ approximates the kernelized Euclidean distance of $\mathbf{x}_i$ and $\mathbf{x}_j$. The retrieval performance of hashing codes will be identical to that of original data points. Except for the binary constraint on $\mathbf{B}$, the other two constraints are called orthogonality constraint and balance constraint respectively. The orthogonality constraint ${{\bf{B}}^\top}{\bf{B}} = n{\bf{I}}$ decorrelates bits of hashing codes. For an extreme example, if the $i$th column and $j$th column are linear correlated or even identical, the retrieval performance will not vary by removing either of them. The balance constraint ${{\bf{B}}^\top}{\bf{1}} = 0$ requires each column of $\mathbf{B}$ has the same number of $1$ and $-1$. For an extreme example, if all elements of the $i$ column of $\mathbf{B}$ is $1$, this column becomes redundant because it does not affect the Hamming distances. Hence, these two constraints were considered to be necessary for good codes~\cite{SH}.\\
\indent Eq.~\eqref{eq:SH} is intractable for large data sets because it requires computing the pairwise distances in the whole data sets to construct the affinity matrix whose element in the $i$th row and $j$th column is ${e^{ - {{\left\| {{{\bf{x}}_i} - {{\bf{x}}_j}} \right\|}^2}/{\sigma ^2}}}$. Furthermore, the binary constraint makes it an NP-hard problem. The authors circumvent these problems by relaxing the binary constraint. The final hashing codes are generated by thresholding eigenfunctions that are designed to avoid computing pairwise distances. On the other hand, anchor graph hashing~\cite{AGH} and discrete graph hashing~\cite{DGH} choose some special points as anchor points. Then, the distances of data points to anchor points are computed to construct a highly sparse affinity matrix so that Eq.~\eqref{eq:SH} can be used for large data sets.\\
\indent ITQ models unimodal hashing as a quantization loss minimization problem:
\begin{equation}\label{eq:ITQ}
\mathop {\arg \min }\limits_{{\bf{B}},{\bf{R}}} \left\| {{\bf{B}} - {\bf{XWR}}} \right\|_F^2,
\end{equation}
where $\mathbf{W}$ is comprised of the first $c$ principal components of $\mathbf{X}$ and $\mathbf{R}$ is an orthogonal matrix. ITQ iteratively computes $\mathbf{B}$ and $\mathbf{R}$ to minimize Eq.~\eqref{eq:ITQ}. In each iteration, $\mathbf{B}$ is thresholded at 0 to generate binary codes. Isotropic hashing (IsoH)~\cite{IsoH} equalizes the importances of principal components. Harmonious hashing~\cite{HH} puts an orthogonality constraint on an auxiliary variable for the code matrix. Unlike ITQ, IsoH and HH that rotate the projected data matrix, ok-means~\cite{okmeans} rotates the code matrix to minimize quantization loss. Despite of principal component analysis (PCA), linear discriminant analysis (LDA) can be used~\cite{LDAHashing}. Neighborhood discriminant hashing (NDH)~\cite{NDH} calculates $\mathbf{W}$ during the minimization procedure rather than pre-computing it by a linear transformation method.\\
\indent All aforementioned unimodal models neglected the balance constraint. Spherical hashing (SpH)~\cite{SpH} and global hashing system (GHS)~\cite{GHS} quantize the distance between a data point and an anchor point. The closer half is denoted as 1, while the further half is denoted 0 or -1. Therefore, the balance constraint can be easily fulfilled. Their major difference is on how to find these anchor points. SpH uses a heuristic algorithm while GHS treats it as a satellite distribution problem of the global positioning system (GPS).\\
\subsection{Multiview Hashing}
Zhang~\emph{et~al.}~\cite{CHMIS-AW} proposed an unsupervised multiview hashing method by extending the unimodal hashing model AGH. It tunes the weights on each view to maximize the performance. Song~\emph{et~al.}~\cite{MFH} jointly consider the local structural information and the relations between local structures to other local structures to design an unsupervised multiview hashing method. Multiview alignment hashing~\cite{MAH} combines the ideas of ok-means and SH for each view. MAH is also an unsupervised multiview hashing method.\\
\indent Multi-graph hashing (MGH)~\cite{MGH} directly combines the graph for the whole data set and the graph for the labeled data to design a semi-supervised multiview hashing method. Semi-supervised multiview discrete hashing (SSMDH)~\cite{SSMDH} predicts the unknown labels by linearly transforming the hashing code matrix of labeled data. It iteratively computes the linearly transformation matrix, hashing code matrix and hashing functions. SSMDH requires the computation of pairwise distances in the whole data set, so it is infeasible for large data set.\\
\subsection{Multimodal Hashing}
Existing multimodal hashing methods can be categorized into supervised and unsupervised ones. Similar to unsupervised unimodal hashing methods, unsupervised multimodal hashing methods aim at preserving the Euclidean distances between each pair of data. Inter-media hashing (IMH)~\cite{mIMH} exploits inter-media consistency and intra-media consistency to generate hashing codes. Like what AGH has done to SH, linear cross-media hashing (LCMH)~\cite{LCMH} uses the distances between each data point and each cluster centroid to construct a sparse affinity matrix. Collective matrix factorization hashing (CMFH)~\cite{CMFH} can be treated as an extension of NDH. For each modality, CMFH consists of two terms: (1) calculating a transformation matrix for the data matrix to match the code matrix, and (2) calculating a transformation matrix for code matrix to match the data matrix.\\
\indent By incorporating label information, supervised multimodal hashing can preserve semantic information and achieve higher accuracy. Cross-modality similarity-sensitive hashing (CMSSH)~\cite{CMSSH} treats hashing as a binary classification problem. Cross-view hashing (CVH)~\cite{CVH} assumes the hashing codes be a linear embedding of the original data points. It extends SH by minimizing the weighted average Hamming distances of hashing codes of training data pairs. The minimization is solved as a generalized eigenvalue problem. The performance of CVH decreases with increasing bit number, because most of the variance is contained in the top few eigenvectors~\cite{MDBE}. Multilatent binary embedding (MLBE)~\cite{MLBE} treats hashing codes as the binary latent factors in the proposed probabilistic model and maps data points from multiple modalities to a common Hamming space. Semantics-preserving hashing (SePH)~\cite{SePH} learns the hashing codes by minimizing the KL-divergence of probability distribution in Hamming space from that in semantic space. CMSSH, MLBE and SePH need to compute the pairwise distances among all data points.\\
\indent Semantic correlation maximization (SCM)~\cite{SCM} circumvents this by learning only one bit each time and the explicit computation of affinity matrix is avoided through several mathematical manipulations. Multi-modal discriminative binary embedding (MDBE)~\cite{MDBE} derives from CMFH. MDBE transform data matrices and label matrix to a latent space and then transform data matrices in latent space to match label matrix.\\
\section{Methodology}
Let us define the used notations first. $\mathbf{X}_i\in\mathcal{R}^{n\times d_i}$ is the data matrix in the $i$th modality, where $n$ is the number of data points and $d_i$ is the dimension of a data point in $i$th modality. Let us assume $\mathbf{X}_i$ be shuffled and zero-centered. $\mathbf{X}_i^l\in\mathcal{R}^{n^l\times d_i}$ is the labeled data matrix comprised of the first $n^l$ rows of $\mathbf{X}$ and $\mathbf{X}_i^u\in\mathcal{R}^{n^u\times d_i}$ is the unlabeled data matrix comprised of the remaining $n^u$ rows of $\mathbf{X}_i$. Hence, we have $n=n^l+n^u$. $\mathbf{B}\in\{0,1\}^{n\times c}$ is the hashing code matrix where $c$ is the code length. $\mathbf{L}^l\in\{-1,1\}^{n_l\times l}$ is the label matrix, where $n_l\leq n$ is the number of labeled data and $l$ is the number of classes. $\mathbf{L}^u\in\mathcal{R}^{n^u\times l}$ is the estimated labels. $\mathbf{p}_i\in\mathcal{R}^{1\times n^u}$ is a vector whose elements are memberships and $\mathbf{P}_i$ is a diagonal matrix whose diagonal elements are $\mathbf{p}_i$. The number of modalities excluding the label modality is $K$. The transformation matrices for estimating labels are $\mathbf{W}_i^L\in\mathcal{R}^{d_i\times l}$ and $\mathbf{V}^L\in\mathcal{R}^{l\times l}$ for the $i$th modality and label matrix, respectively. The transformation matrices for generating hashing codes are $\mathbf{W}_i^B\in\mathcal{R}^{d_i\times c}$ and $\mathbf{V}^B\in\mathcal{R}^{l\times c}$, respectively.
The superscript $l$ is short for ``labeled'', while $u$ is short for ``unlabeled''. They indicate the correspondence to labeled or unlabeled data. The superscript $L$ and $B$ stands for ``generating label matrix'' and ``generating binary code matrix'', respectively.\\
\indent As illustrated in Fig.~\ref{fig:scheme}, the core parts of the proposed semi-supervised multimodal hashing (SSMH) are supervised hashing for generating labels, label estimation and supervised hashing for generating hashing codes which are highlighted by green blocks. We introduces them consecutively in the following subsections.\\
\subsection{Supervised hashing for estimating labels}
\label{subsec:hashingL}
By treating label matrix as a special modality, we formulate the supervised hashing methods for estimating labels as:
\begin{equation}\label{eq:premodel}
\begin{array}{l}
\mathop {\arg \min }\limits_{{\bf{H}}^l,{{\bf{W}}_i}} E=\frac{1}{2}\sum\limits_{i = 1}^K {\left\| {{\bf{H}}^l - {\bf{X}}_i^l{\bf{W}}_i^L} \right\|_F^2}  + \frac{\alpha}{2} \left\| {{\bf{H}}^l - {\bf{L}}{{\bf{V}}^L}} \right\|_F^2\\
\;\;\;\;\;\;\;\;\;\;\;\;\;\;\;\;\;\;\;\;\;\;\;\;\;s.t.\;\;\;\;{\bf{H}}^l \in {\left\{ { - 1,1} \right\}^{{n^l} \times l}}
\end{array}
\end{equation}
where ${\bf{H}}^l$ is a temporary variable and $\alpha$ is a pre-defined real positive constant. Eq.~\eqref{eq:premodel} is generally ill-posed since we have K+2 unknown variables and K+1 known constant matrices. Hence, it should be regularized. Inspired by the orthogonality regularization proposed in~\cite{aaaiorth}, we can modify Eq.~\eqref{eq:premodel} as:
\begin{equation}
\begin{array}{l}\label{eq:modelL}
\mathop {\arg \min }\limits_{{\bf{H}}^l,{{\bf{W}}_i}} E=\frac{1}{2}\sum\limits_{i = 1}^K {\left\| {{\bf{H}}^l - {\bf{X}}_i^l{\bf{W}}_i^L} \right\|_F^2}  + \frac{\alpha}{2}\left\| {{\bf{H}}^l - {\bf{L}}{{\bf{V}}^L}} \right\|_F^2 + \\
\;\;\;\;\;\;\;\;\;\;\frac{\gamma}{2} \left( {\sum\limits_{i = 1}^K {\left\| {{{\bf{W}}_i^L}^\top{\bf{W}}_i^L - {\bf{I}}} \right\|_F^2}  + \left\| {{{\bf{V}}^L}^\top{{\bf{V}}^L} - {\bf{I}}} \right\|_F^2} \right)\\
\;\;\;\;\;\;\;\;\;\;\;\;\;\;\;\;\;\;\;\;\;\;\;\;\;s.t.\;\;\;\;{\bf{H}}^l \in {\left\{ { - 1,1} \right\}^{{n^l} \times l}}
\end{array}
\end{equation}
where $\mathbf{I}$ is the identity matrix and $\gamma$ is a pre-defined real positive constant.\\
\indent Eq.~\eqref{eq:modelL} is solved by iteratively calculating $\mathbf{H}^l$, $\mathbf{W}_i^L$ and $\mathbf{V}^L$. Take the first derivative with respect to $\mathbf{W}_i^L$:
\begin{equation}\label{eq:WL}
\frac{\partial E}{\partial \mathbf{W}_i^L}={\mathbf{X}_i^l}^\top\left(\mathbf{X}_i^l\mathbf{W}_i^L-\mathbf{H}^l\right)+\gamma\mathbf{W}_i^L\left({\mathbf{W}_i^L}^\top\mathbf{W}_i^L-\mathbf{I}\right).
\end{equation}
Similarly, we have
\begin{equation}\label{eq:VL}
\frac{\partial E}{\partial \mathbf{V}^L}=\alpha{\mathbf{L}^l}^\top\left(\mathbf{L}^l\mathbf{V}^L-\mathbf{H}^l\right)+\gamma\mathbf{V}^L\left({\mathbf{V}^L}^\top\mathbf{V}^L-\mathbf{I}\right).
\end{equation}
Taking the first derivative of Eq.~\eqref{eq:modelL} with respect to $\bf{H}$ and set it as 0, we have
\begin{equation}\label{eq:BL}
\mathbf{H}^l=\textrm{sign}\left(\sum_{i=1}^K{\mathbf{X}_i^l\mathbf{W}_i^L}+\alpha\mathbf{L}^l\mathbf{V}^L\right).
\end{equation}
Gradient descent method is used for updating $\mathbf{W}_i^L$ and $\mathbf{V}^L$. The optimization procedure is shown in \textbf{Algorithm 1}, where $\Delta t$ is the step size. The first terms of derivatives with respect to $\mathbf{W}_i^L$ and $\mathbf{V}^L$ are normalized so that we can fix $\Delta t$ for all our experiments, otherwise $\Delta t$ should be tuned for different size of $\mathbf{X}_i^l$ because ${\mathbf{X}_i^l}^\top\mathbf{X}_i^l$ increases dramatically with large $\mathbf{X}_i^l$.\\
\begin{algorithm}
\caption{Supervised Hashing for Generating Labels}
\label{alg:1}
\begin{algorithmic}[1]
\Require $\alpha$, $\gamma$, $\Delta t$, $\mathbf{X}_i^l$, $\mathbf{L}^l$
\State Initialize $\mathbf{W}_i^L$, $\mathbf{V}^L$
\While {$E$ not converged}
\State Update $\mathbf{H}^l$ using Eq.~\eqref{eq:BL}.
\State $\mathbf{V}^L\leftarrow\mathbf{L}^L-\Delta t\cdot{\partial E}/{\partial \mathbf{V}^L}$ using Eq.~\eqref{eq:VL}
\State $\mathbf{W}_i^L\leftarrow\mathbf{W}_i^L-\Delta t\cdot{\partial E}/{\partial \mathbf{W}_i^L}$ using Eq.~\eqref{eq:WL}
\EndWhile
\Ensure $\mathbf{W}_i^L$,$\mathbf{V}_i^L$
\end{algorithmic}
\end{algorithm}
\subsection{Generating labels}
We generate $\mathbf{H}^u_i$ by $\mathbf{W}_i^L$ for the unlabeled data in the $i$th modality. Unlike $\mathbf{H}^l$ which is a unified code matrix for all modalities, $\mathbf{H}^u_i$ generally differs from each other. We assume $\mathbf{V}^L$ can transform the unknown label matrix $\mathbf{L}^u$ to match $\mathbf{H}^u_i$ in a certain probability. The probability that $\mathbf{L}^u$ transformed by $\mathbf{V}^L$ matches $\mathbf{H}^u_i$ is denoted as $\mathbf{p}_i\in(0,1)^{1\times n^u}$. The $r$th element of $\mathbf{p}_i$ corresponds to that probability of the $r$th label vector, i.e., the $r$th row of $\mathbf{L}^u$.\\
\indent In fuzzy c-means clustering (FCM), the membership indicates the probability that a data point belongs to a centroid. This probability is determined by the distance between the data point and the centroid. Similar to FCM, we use the following object function to estimate labels,
\begin{equation}\label{eq:temp}
\begin{array}{c}
\mathop {\arg \min }\limits_{{{\bf{L}}^u},\mathbf{p}_i} E = \sum\limits_{i = 1}^K {\sum\limits_{r = 1}^{{n^u}} {{{\bf{p}}_i^m}\left( r \right)\left\| {{{\bf{L}}^u}\left( {r,:} \right){{\bf{V}}^L} - {\bf{H}}_i^u} \right\|_F^2} },\\
s.t.\;\;\sum_{i=1}^K{\mathbf{p}_i}=\mathbf{1}, \mathbf{p}_i\geq\mathbf{0}
\end{array}
\end{equation}
where $m$ is the fuzzier that determines the level of fuzziness of $\mathbf{H}_i^u$, $\mathbf{L}^u(r,:)$ is the $r$th row of $\mathbf{L}^u$, $\mathbf{p}_i(r)$ is the $r$th element of $\mathbf{p}_i$, $\mathbf{1}$ is an $1\times n^u$ vector of ones, $\mathbf{0}$ is an $1\times n^u$ vector of zeros and
\begin{equation}
\mathbf{H}_i^u=\textrm{sign}\left(\mathbf{X}^u_i\mathbf{W}_i^L\right).
\end{equation}
\indent Eq.~\eqref{eq:temp} can be written in matrix form:
\begin{equation}\label{eq:fuzzy}
\mathop {\arg \min }\limits_{{{\bf{L}}^u},\mathbf{P}_i^m} E = \sum\limits_{i = 1}^K {tr\left( {\left( {{{\bf{L}}^u}{{\bf{V}}^L} - {\bf{H}}_i^u} \right)\left( {{{\bf{L}}^u}{{\bf{V}}^L} - {\bf{H}}_i^u} \right)^\top{{\bf{P}}_i^m}} \right)}
\end{equation}
where $tr()$ means the trace of a matrix and $\mathbf{P}_i$ is a diagonal matrix of which the diagonal elements are $\mathbf{p}_i$. Similar to FCM, we iteratively minimize Eq.~\eqref{eq:fuzzy}. Taking partial derivative with respect to $\mathbf{L}^u$ and setting it to 0, we have
\begin{equation}\label{eq:updateL}
\mathbf{L}^u = \left(\sum_{i=1}^K{\mathbf{P}_i^m\mathbf{X}_i^u}\right){\mathbf{V}^L}^\top\left(\mathbf{V}^L{\mathbf{V}^L}^\top\right)^{-1},
\end{equation}
where the superscript ``-1'' means inverse matrix. $\mathbf{V}^L$ is a square matrix, so is $\mathbf{V}^L{\mathbf{V}^L}^\top$. When $\mathbf{V}^L$ is randomly initialized and hence it is generally invertible, we empirically found that the hashing method in Subsection~\ref{subsec:hashingL} led to an invertible $\mathbf{V}^L$, too. It is easy to figure out if $\mathbf{V}^L$ is invertible, $\mathbf{V}^L{\mathbf{V}^L}^\top$ is also invertible and its inverse matrix is $({\mathbf{V}^L}^{-1})^\top{\mathbf{V}^L}^{-1}$.\\
\indent By introducing Lagrangian multiplier $\lambda$, Eq.~\eqref{eq:fuzzy} can be formulated as an unconstrained minimization problem:
\begin{equation}
\begin{array}{c}
\mathop {\arg \min }\limits_{{{\bf{L}}^u},\mathbf{P}_i^m} E = \sum\limits_{i = 1}^K {tr\left( {\left( {{{\bf{L}}^u}{{\bf{V}}^L} - {\bf{H}}_i^u} \right)\left( {{{\bf{L}}^u}{{\bf{V}}^L} - {\bf{H}}_i^u} \right)^\top{{\bf{P}}_i^m}} \right)}\\
+\lambda\left(\sum_{i=1}^K{\mathbf{P}_i}-\mathbf{I}\right).
\end{array}
\end{equation}
Let us define
\begin{equation}
\mathbf{D}_i = \left( {{{\bf{L}}^u}{{\bf{V}}^L} - {\bf{H}}_i^u} \right)^\top.
\end{equation}
It can be deduced that
\begin{equation} \label{eq:updateP}
\mathbf{P}_i = \frac{1}{\sum_{k=1}^K{\left(\frac{\left(\mathbf{D}_i^\top\mathbf{D}_i\right)\circ \mathbf{I}}{\left(\mathbf{D}_k^\top\mathbf{D}_k\right)\circ\mathbf{I}}\right)^\frac{1}{m-1}}}
\end{equation}
where the division and exponent of matrices are element-wise. In Eq.~\eqref{eq:updateP}, it is unnecessary to compute $\mathbf{D}_i^\top\mathbf{D}_i$. Only the diagonal elements are non-zeros after the element-wise multiplication with $\mathbf{I}$. Hence, we just compute the squared Frobenius norm of each row of $\mathbf{D}_i$. The key steps are summarized in \textbf{Algorithm 2}.\\
\begin{algorithm}
\caption{Algorithm for generating labels}
\label{alg:1}
\begin{algorithmic}[1]
\Require $\mathbf{W}_i^L$, $\mathbf{V}^L$, $\mathbf{X}_i^u$
\State Initialize $\mathbf{P}_i$ and $\mathbf{L}^u$.
\While {maximum iteration number not reached}
\State Update $\mathbf{P}_i$ using Eq.~\eqref{eq:updateP}.
\State Update $\mathbf{L}^u$ using Eq.~\eqref{eq:updateL}.
\EndWhile
\Ensure $\mathbf{L}^u$
\end{algorithmic}
\end{algorithm}
\subsection{Supervised hashing for generating hashing codes}
The hashing method proposed in this subsection is slightly different from that introduced in Subsection~\ref{subsec:hashingL}. As the estimated labels are not actual labels, we should deduce their effects on calculating the transformation matrix. Let us define $\mathbf{B}^l$ and $\mathbf{B}^u$ as the hashing code matrices for labeled and unlabeled data, respectively. We modify Eq.~\eqref{eq:modelL} to
\begin{equation}
\begin{array}{c}
\mathop {\arg \min }\limits_{{\bf{B}},{{\bf{W}}_i^B},\mathbf{V}^B} E=\\
\beta^l\left(\frac{1}{2}\sum\limits_{i = 1}^K {\left\| {{\bf{B}}^l - {\bf{X}}_i^l{\bf{W}}_i^B} \right\|_F^2}  + \frac{\alpha}{2}\left\| {{\bf{B}}^l - {\bf{L}}^l{{\bf{V}}^B}} \right\|_F^2\right) + \\
\beta^u\left(\frac{1}{2}\sum\limits_{i = 1}^K {\left\| {{\bf{B}}^u - {\bf{X}}_i^l{\bf{W}}_i^B} \right\|_F^2}  + \frac{\alpha}{2} \left\| {{\bf{B}}^u - {\bf{L}}^u{{\bf{V}}^B}} \right\|_F^2\right)+\\
\frac{\gamma}{2} \left( {\sum\limits_{i = 1}^K {\left\| {{{\bf{W}}_i^B}^\top{\bf{W}}_i^B - {\bf{I}}} \right\|_F^2}  + \left\| {{{\bf{V}}^B}^\top{{\bf{V}}^B} - {\bf{I}}} \right\|_F^2} \right)\\
s.t.\;\;\;\;{\bf{B}} \in {\left\{ { - 1,1} \right\}^{{n} \times l}}
\end{array}
\end{equation}
where $\beta^l$ and $\beta^u$ is pre-defined positive constant and $\mathbf{B}$ is the concatenation matrix of $\mathbf{B}^l$ and $\mathbf{B}^u$. The minimization procedure is similar to \textbf{Algorithm 1}. Therefore, to save space, we only give the substitutes of Eq.~\eqref{eq:BL}, Eq.~\eqref{eq:VL} and Eq.~\eqref{eq:WL}. To generate hashing codes, Eq.~\eqref{eq:WL} should be substituted by
\begin{equation}
\begin{array}{c}
\frac{\partial E}{\partial \mathbf{W}_i^B}=\beta^l{\mathbf{X}_i^l}^\top\left(\mathbf{X}_i^l\mathbf{W}_i^B-\mathbf{B}^l\right)+\beta^u{\mathbf{X}_i^u}^\top\left(\mathbf{X}_i^u\mathbf{W}_i^B-\mathbf{B}^u\right)\\
+2\gamma\mathbf{W}_i^B\left({\mathbf{W}_i^B}^\top\mathbf{W}_i^B-\mathbf{I}\right),
\end{array}
\end{equation}
Eq.~\eqref{eq:VL} should be substituted by
\begin{equation}
\begin{array}{c}
\frac{\partial E}{\partial \mathbf{V}^B}=\beta^l\alpha{\mathbf{L}^l}^\top\left(\mathbf{L}^l\mathbf{V}^B-\mathbf{B}^l\right)+\alpha\beta^u{\mathbf{L}^u}^\top\left(\mathbf{L}^u\mathbf{V}^B-\mathbf{B}^u\right)\\
+\gamma\mathbf{V}^B\left({\mathbf{V}^B}^\top\mathbf{V}^B-\mathbf{I}\right),
\end{array}
\end{equation}
and Eq.~\eqref{eq:BL} should be substituted by
\begin{equation}
\left\{ {\begin{array}{*{20}{c}}
{{{\bf{B}}^{\rm{l}}} = \textrm{sign}\left(\sum\limits_{i = 1}^K {{\bf{X}}_i^l{\bf{W}}_i^B + \alpha {{\bf{L}}^u}{{\bf{V}}^B}}\right) }\\
{{{\bf{B}}^u} = \textrm{sign}\left(\sum\limits_{i = 1}^K {{\bf{X}}_i^u{\bf{W}}_i^B + \alpha {{\bf{L}}^u}{{\bf{V}}^B}}\right) }
\end{array}} \right. .
\end{equation}
\subsection{Implementation details}
\textbf{Initialization}. In our experiment, all transformation matrices, including $\mathbf{W}_i^L$, $\mathbf{W}_i^B$, $\mathbf{V}^L$ and $\mathbf{V}^B$, were randomly initialized and normalized. In the label generation step, $\mathbf{L}^u$ was initialized as an arbitrary $\mathbf{H}^u$.\\
\indent\textbf{Parameter setting}. We set $\alpha=100$ and $\gamma=0.01$ for two supervised hashing methods. $\beta^l$ was set as $n^u/n$ and $\beta^u$ was set as $0.1n^l/n$. $\Delta t$ was set as 0.001. The maximum iteration number for two supervised hashing method was set as 400, while it was set as 15 for label generation method.\\
\section{Experimental Results}
\label{sec:experiment}
\subsection{Data sets and baselines}
\textbf{MIRFlickr}~\cite{MIRFlickr} contains 25,000 entries each of which consists of 1 image, several textual tags and labels. Following literature~\cite{SePH}, we only keep those textural tags appearing at least 20 times and remove entries which have no label. Hence, 20,015 entries are left. For each entry, the image is represented by a 512-dimensional GIST~\cite{GIST} descriptors and the text is represented by a 500-dimensional feature vector derived from PCA on index vectors of the textural tags. 5\% entries are randomly selected for testing and the remaining entries are used as training set. In the training set, we use 10\%, 50\% and 90\% labels to construct three partially labeled data sets. Ground-truth semantic neighbors for a test entry, i.e, a query, are defined as those sharing at least one label. \\
\indent The proposed method was compared with five state-of-the-art supervised multimodal hashing methods CMSSH~\cite{CMSSH}, CVH~\cite{CVH}, SCM~\cite{SCM}, SePH~\cite{SePH} and MDBE~\cite{MDBE} and one semi-supervised method S3FH~\cite{S3FH}. For supervised methods, 100\% labels are used to train their models.
\subsection{Results and analysis}
Mean average precision (MAP) which varies between 0 and 1 is a widely-used evaluation metric for retrieval performance. Table~\ref{tb:map} shows the MAP of compared methods. ``Image-text'' means using images searching texts, while ``text-image'' means using texts searching images. In Table~\ref{tb:map}, the performances are sorted in ascendant order according to the criteria that if method I performs better on at least three experiments than method II, then method I is supposed to be superior. It can be seen that with only 10\% labels, SSMH approximates the worst full-supervised methods. With 50\% labels, SSMH gets a medium performance among compared methods. With 90\% labels, SSMH surpasses all compared methods except for MDBE. However, the results of SSMH(90\%) and MDBE does not differ significantly from each other.
\begin{table*}[!h]
\caption{MAP results on MIRFlickr data sets.}
\label{tb:map}
\begin{center}
\begin{tabular}{|c|c|c	c	c	c	c|}
\hline
\multirow{ 2}{*}{Task} & \multirow{2}{*}{Method} & \multicolumn{5}{|c|}{\bf MIRFlickr} \\ \cline{3-7} & & 16 bits & 32 bits & 64 bits & 96 bits & 128 bits\\ \hline
\multirow{11}{*}{Image-Text}&S3FH(10\%)& 0.5894&0.5902&0.5951&0.5947&0.5912\\
&CMSSH & 0.5966 & 0.5674 & 0.5581 & 0.5692 & 0.5701 \\ 
&SSMH(10\%)&0.5802&0.5945&0.5953&0.5956&0.5916\\
&S3FH(50\%)&0.5954&0.6028&0.6057&0.6044&0.6099\\
&CVH &0.6591 & 0.6145 & 0.6133 & 0.6091 & 0.6052\\ 
&S3FH(90\%)&0.6116&0.6273&0.6145&0.6125&0.6191\\
&SSMH(50\%)&0.6326&0.6333&0.6372&0.6381&0.6344\\
&SCM &0.6251 &0.6361 &0.6417 &0.6446 &0.6480\\ 
&SePH &0.6505 &0.6447 &0.6453 &0.6497 &0.6612\\ 
&SSCM(90\%)&0.6612&0.6654&0.6818&0.6906&0.6950\\
&MDBE&0.6784&0.6950&0.6983&0.7048&0.7056\\
\hline
\multirow{11}{*}{Text-Image}&S3FH(10\%)&0.5743&0.5835&0.5976&0.5997&0.5903\\
&SSMH(10\%)&0.5891&0.5994&0.6008&0.5924&0.5982\\ 
&CVH& 0.6495 & 0.6213 & 0.6179 & 0.6050 & 0.5948\\ 
&S3FH(50\%)&0.5892&0.6069&0.6120&0.6126&0.6161\\
&S3FH(90\%)&0.6281&0.6299&0.6315&0.6350&0.6329\\
&SCM& 0.6194 & 0.6302 & 0.6377 & 0.6377 & 0.6417\\ 
&SSMH(50\%)&0.6318&0.6335&0.6402&0.6411&0.6395\\
&CMSSH &0.6613 &0.6510 & 0.6756 & 0.6643 & 0.6471\\
&SePH& 0.6745 &0.6824 & 0.6917 &0.7059 &0.7110\\ 
&SSCM(90\%)&0.6672&0.7146&0.7254&0.7255&0.7332\\
&MDBE&0.6723&0.7237&0.7353&0.7355&0.7387\\
\hline
\end{tabular}
\end{center}
\end{table*}
\section{Conclusion}
In this paper, we proposed a semi-supervised multimodal hashing method (SSMH). SSMH first utilizes a supervised hashing method to generate a code matrix of the same dimension as label matrix for labeled data. Then, it transformed the unlabel data matrices to generate candidate label matrices. In each modality, a membership variable was introduced to represent the probability that the transformed label matrix for unlabeled data belongs to this modality. By iteratively calculating the membership variables and estimating label matrix, SSMH generated a label matrix for unlabeled data. Finally, the supervised hashing method in the first step was modified to generate a unified hashing code matrix. Experiments shew that the performance of SSMH approximately ranged within that of the worst compared supervised method and that of the best one, given that the percentage of available labels ranged within 10\%-90\%.

\bibliographystyle{IEEEtran}
\bibliography{Output4}

\begin{thebibliography}{10}
\providecommand{\url}[1]{#1}
\csname url@samestyle\endcsname
\providecommand{\newblock}{\relax}
\providecommand{\bibinfo}[2]{#2}
\providecommand{\BIBentrySTDinterwordspacing}{\spaceskip=0pt\relax}
\providecommand{\BIBentryALTinterwordstretchfactor}{4}
\providecommand{\BIBentryALTinterwordspacing}{\spaceskip=\fontdimen2\font plus
\BIBentryALTinterwordstretchfactor\fontdimen3\font minus
  \fontdimen4\font\relax}
\providecommand{\BIBforeignlanguage}[2]{{%
\expandafter\ifx\csname l@#1\endcsname\relax
\typeout{** WARNING: IEEEtran.bst: No hyphenation pattern has been}%
\typeout{** loaded for the language `#1'. Using the pattern for}%
\typeout{** the default language instead.}%
\else
\language=\csname l@#1\endcsname
\fi
#2}}
\providecommand{\BIBdecl}{\relax}
\BIBdecl

\bibitem{S3FH}
J.~Wang, G.~Li, P.~Pan, and X.~Zhao, ``Semi-supervised semantic factorization
  hashing for fast cross-modal retrieval,'' \emph{Multimedia Tools and
  Applications}, vol.~76, no.~19, pp. 20\,197--20\,215, Oct 2017.

\bibitem{SH}
Y.~Weiss, A.~Torralba, and R.~Fergus, ``Spectral hashing,'' in \emph{Advances
  in Neural Information Processing Systems}, 2008, pp. 1753--1760.

\bibitem{AGH}
W.~Liu, J.~Wang, and S.-f. Chang, ``Hashing with graphs,'' in
  \emph{International Conference on Machine Learning}, 2011.

\bibitem{DGH}
W.~Liu, C.~Mu, S.~Kumar, and S.-F. Chang, ``Discrete graph hashing,'' in
  \emph{Advances in Neural Information Processing Systems}, 2014.

\bibitem{IsoH}
W.~Kong and W.-J. Li, ``Isotropic hashing,'' in \emph{Advances in Neural
  Information Processing Systems}, 2012, pp. 1646--1654.

\bibitem{HH}
B.~Xu, J.~Bu, Y.~Lin, C.~Chen, X.~He, and D.~Cai, ``Harmonious hashing,'' in
  \emph{International Joint Conference on Artificial Intelligence}, 2013, pp.
  1820--1826.

\bibitem{okmeans}
M.~Norouzi and D.~J. Fleet, ``Cartesian k-means,'' in \emph{IEEE Conference on
  Computer Vision and Pattern Recognition}, 2013, pp. 3017--3024.

\bibitem{LDAHashing}
C.~Strecha, A.~M. Bronstein, M.~M. Bronstein, and P.~Fua, ``{LDAH}ash: Improved
  matching with smaller descriptors,'' \emph{IEEE Transactions on Pattern
  Analysis and Machine Intelligence}, vol.~34, no.~1, pp. 66--78, May 2012.

\bibitem{NDH}
J.~Tang, Z.~Li, M.~Wang, and R.~Zhao, ``Neighborhood discriminant hashing for
  large-scale image retrieval,'' \emph{IEEE Transactions on Image Processing},
  vol.~24, no.~9, pp. 2827--2840, Sept 2015.

\bibitem{SpH}
H.~Jae-Pil, L.~Youngwoon, H.~Junfeng, C.~Shih-Fu, and Y.~Sung-Eui, ``Spherical
  hashing,'' in \emph{IEEE Conference on Computer Vision and Pattern
  Recognition}, 2012, pp. 2957--2964.

\bibitem{GHS}
D.~Tian and D.~Tao, ``Global hashing system for fast image search,'' \emph{IEEE
  Transactions on Image Processing}, vol.~26, no.~1, pp. 79--89, Jan 2017.

\bibitem{CHMIS-AW}
D.~Zhang, F.~Wang, and L.~Si, ``Composite hashing with multiple information
  sources,'' in \emph{In Proceedings of the International ACM SIGIR Conference
  on Research and Development in Information Retrieval}, 2011, pp. 225--234.

\bibitem{MFH}
J.~Song, Y.~Yang, Z.~Huang, H.~T. Shen, and J.~Luo, ``Effective multiple
  feature hashing for large-scale near-duplicate video retrieval,''
  \emph{Trans. Multi.}, vol.~15, no.~8, pp. 1997--2008, Dec. 2013.

\bibitem{MAH}
L.~Liu, M.~Yu, and L.~Shao, ``Multiview alignment hashing for efficient image
  search,'' \emph{IEEE Transactions on Image Processing}, vol.~24, no.~3, pp.
  956--966, March 2015.

\bibitem{MGH}
J.~Cheng, C.~Leng, P.~Li, M.~Wang, and H.~Lu, ``Semi-supervised multi-graph
  hashing for scalable similarity search,'' \emph{Computer Vision and Image
  Understanding}, vol. 124, no. Supplement C, pp. 12 -- 21, 2014.

\bibitem{SSMDH}
C.~Zhang and W.~S. Zheng, ``Semi-supervised multi-view discrete hashing for
  fast image search,'' \emph{IEEE Transactions on Image Processing}, vol.~26,
  no.~6, pp. 2604--2617, June 2017.

\bibitem{mIMH}
J.~Song, Y.~Yang, Y.~Yang, Z.~Huang, and H.~T. Shen, ``Inter-media hashing for
  large-scale retrieval from heterogeneous data sources,'' in \emph{Proceedings
  of ACM SIGMOD International Conference on Management of Data}, 2013, pp.
  785--796.

\bibitem{LCMH}
X.~Zhu, Z.~Huang, H.~T. Shen, and X.~Zhao, ``Linear cross-modal hashing for
  efficient multimedia search,'' in \emph{Proceedings of ACM International
  Conference on Multimedia}, 2013, pp. 143--152.

\bibitem{CMFH}
G.~Ding, Y.~Guo, and J.~Zhou, ``Collective matrix factorization hashing for
  multimodal data,'' in \emph{IEEE Conference on Computer Vision and Pattern
  Recognition}, 2014, pp. 2083--2090.

\bibitem{CMSSH}
M.~M. Bronstein, A.~M. Bronstein, F.~Michel, and N.~Paragios, ``Data fusion
  through cross-modality metric learning using similarity-sensitive hashing,''
  in \emph{IEEE Conference on Computer Vision and Pattern Recognition}, 2010,
  pp. 3594--3601.

\bibitem{CVH}
S.~Kumar and R.~Udupa, ``Learning hash functions for cross-view similarity
  search,'' in \emph{Proceedings of International Joint Conference on
  Artificial Intelligence}, July 2011.

\bibitem{MDBE}
D.~Wang, X.~Gao, X.~Wang, L.~He, and B.~Yuan, ``Multimodal discriminative
  binary embedding for large-scale cross-modal retrieval,'' \emph{IEEE
  Transactions on Image Processing}, vol.~25, no.~10, pp. 4540--4554, Oct.
  2016.

\bibitem{MLBE}
Y.~Zhen and D.-Y. Yeung, ``A probabilistic model for multimodal hash function
  learning.'' in \emph{KDD}.\hskip 1em plus 0.5em minus 0.4em\relax ACM, 2012,
  pp. 940--948.

\bibitem{SePH}
Z.~Lin, G.~Ding, J.~Han, and J.~Wang, ``Cross-view retrieval via
  probability-based semantics-preserving hashing,'' \emph{IEEE Transactions on
  Cybernetics}, vol.~47, no.~12, pp. 4342--4355, Dec 2017.

\bibitem{SCM}
D.~Zhang and W.-J. Li, ``Large-scale supervised multimodal hashing with
  semantic correlation maximization,'' in \emph{Proceedings of the
  Twenty-Eighth AAAI Conference on Artificial Intelligence}, 2014, pp.
  2177--2183.

\bibitem{aaaiorth}
D.~Wang, P.~Cui, M.~Ou, and W.~Zhu, ``Deep multimodal hashing with orthogonal
  regularization,'' in \emph{Proceedings of International Joint Conference on
  Artificial Intelligence}, 2015, pp. 2291-- 2297.

\bibitem{MIRFlickr}
M.~J. Huiskes and M.~S. Lew, ``The {MIR} flickr retrieval evaluation,'' in
  \emph{Proceedings of the ACM International Conference on Multimedia
  Information Retrieval}, 2008.

\bibitem{GIST}
A.~Oliva and A.~Torralba, ``Modeling the shape of the scene: A holistic
  representation of the spatial envelope,'' \emph{International Journal of
  Computer Vision}, vol.~42, no.~3, pp. 145--175, May 2001.

\end{thebibliography}

\end{document}